\pdfoutput=1

\documentclass[twocolumn,prd,aps,superscriptaddress
,showpacs,
nofootinbib,preprintnumbers,floats,floatfix]{revtex4}

\usepackage{graphicx}
\usepackage{amssymb,amsmath,amsfonts,amsthm}
\usepackage{amssymb}
\usepackage{epstopdf}
\usepackage{slashed}
\usepackage[colorlinks=true]{hyperref}
\hypersetup{citecolor=black,linkcolor=black}

\newcommand{\f}{\begin{equation}}
\newcommand{\ff}{\end{equation}}

\newcommand{\bea}{\begin{eqnarray}}
\newcommand{\eea}{\end{eqnarray}}
\newcommand{\be}{\begin{equation}}
\newcommand{\bse}{\begin{subequations}}
\newcommand{\ese}{\end{subequations}}
\newcommand{\bear}{\begin{eqnarray}}
\newcommand{\eear}{\end{eqnarray}}
\newcommand{\ba}{\begin{array}}
\newcommand{\ea}{\end{array}}
\newcommand{\ee}{\end{equation}}

\def\fillbox#1{\hbox to #1{\vbox to #1{\vfil}\hfil}}

\begin{document}

\title{A Zero-Parameter Extension of General Relativity with Varying Cosmological Constant}

\author{Stephon Alexander}
\affiliation{Department of Physics, Brown University, Providence, RI, 02906, USA}

\author{Marina Cort\^{e}s}
\affiliation{Perimeter Institute for Theoretical Physics, 31 Caroline Street North,  Waterloo, Ontario N2J 2Y5, Canada}
\affiliation{Instituto de Astrof\'{\i}sica e Ci\^{e}ncias do Espa\c{c}o,
Faculdade de Ci\^{e}ncias, Universidade de Lisboa, 1769-016 Lisboa, Portugal}
\author{Andrew R. Liddle}
\affiliation{Perimeter Institute for Theoretical Physics, 31 Caroline Street North,  Waterloo, Ontario N2J 2Y5, Canada}
\affiliation{Instituto de Astrof\'{\i}sica e Ci\^{e}ncias do Espa\c{c}o,
Faculdade de Ci\^{e}ncias, Universidade de Lisboa, 1769-016 Lisboa, Portugal}
\author{Jo\~{a}o Magueijo}
\affiliation{Theoretical Physics Group, The Blackett Laboratory, Imperial College, Prince Consort Rd., London, SW7 2BZ, United Kingdom}
\author{Robert Sims}
\affiliation{Department of Physics, Brown University, Providence, RI, 02906, USA}
\author{Lee Smolin}
\affiliation{Perimeter Institute for Theoretical Physics, 31 Caroline Street North,  Waterloo, Ontario N2J 2Y5, Canada}

\date{\today}

\begin{abstract}
We provide a new extension of general relativity (GR) which has the remarkable property of being more constrained than GR plus a cosmological constant, having one less free parameter.  This is implemented by allowing the cosmological constant to have a consistent space-time variation, through coding  its dynamics in the torsion tensor.  We demonstrate this mechanism by adding a `quasi-topological' term to the Einstein action, which naturally realizes a dynamical torsion with an automatic satisfaction of the Bianchi identities. Moreover, variation of the action with respect to this dynamical $\Lambda$ fixes it in terms of other variables, thus providing a scenario with \emph{less} freedom than general relativity with a cosmological constant. Once matter is introduced, at least in the homogeneous and isotropic reduction, $\Lambda$ is uniquely determined by the field content of the model. We make an explicit construction using the Palatini formulation of GR and describe the striking properties of this new theory. We also highlight some possible extensions to the theory. A companion paper \cite{FRW1} explores the Friedmann--Robertson--Walker reduction for cosmology, and future work will study Solar System tests of the theory.
\end{abstract}


\maketitle

\section{Introduction}

Given the number of unresolved problems facing the standard model of cosmology (coincidence, graceful exit, dark matter, etc.), researchers have constructed many extensions of the standard model that are equipped with new parameters or functions.  With the proliferation of such candidate parameters, we are often faced with both observational degeneracies and an increased need to fine-tune parameters.  A challenge is to instead enhance the standard model while  {\em decreasing} the number of free parameters. In this paper we introduce a theory of gravity which has no free parameters apart from the gravitational constant, and hence is more constrained than general relativity (GR) with a cosmological constant. 

The new theory of gravity we describe has remarkable properties. Motivated by a
new principle of {\it quasi-topological dynamics}, together with an approximate duality symmetry, we add a uniquely-determined term to the gravitational action. This additional term permits the cosmological `constant' $\Lambda$ to become a variable without violating the consistency conditions based on the
Bianchi identities, and without the need to insert a kinetic term. Varying the action with respect to this dynamical $\Lambda$ then fixes it in terms of the other fields, while still allowing it to vary with space-time position.

This is not what is done in general relativity. There, by taking the covariant divergence of the Einstein equations, one sees that the Bianchi identities force $\Lambda$ to be constant.
Additionally, one  cannot vary the action with respect to $\Lambda$, as this yields only physically-useless solutions where the space-time volume $\sqrt{-g}$ is forced to be zero.\footnote{By contrast, it appears one can self-consistently choose to vary with respect to the other constant in the Einstein--Hilbert action, the gravitational coupling strength, as recently proposed by Lombriser \cite{Lombriser}. This procedure also gives a scenario substantially different from GR, that bears some resemblance to the `sequestering' scenarios \cite{KP,LS-GL}.}
Hence in GR one is required to leave $\Lambda$ as a free parameter that can be fixed only via observation. Generating dark energy dynamics within GR requires, at its simplest, the introduction of new `quintessence' degrees of freedom.
 
This introduces new parameters and fields, which normally supply an abundance of additional free parameters \cite{quint-review,modifiedg-review}. While the cosmological constant continues to give an excellent account of all reliable data, there is a sense of dissatisfaction unless some attempt is made to compare with alternatives. Unfortunately, these alternatives are essentially all motivated solely by the desire to have some alternative to compare with the pure cosmological constant model. In doing so, they introduce numerous possible new functional degrees of freedom, at best tenuously linked to other ideas in fundamental physics. 

Within general relativity we cannot give $\Lambda$ an explicit space-time dependence because that violates the Bianchi identity, though phenomenological variations are still sometimes written in by hand \cite{OC}. Quintessence scenarios avoid this by making $\Lambda$ a function of a field $\phi$ and giving that field dynamics by introducing a kinetic term, unmotivated by anything other than matching observations and additionally making the equations second order. So quintessence requires three additions: 1)  a new variable, $\phi$, 2) a kinetic term associated with it, and 3) a free function which is the potential for that variable. And this is the simplest conventional route to dark energy dynamics; many proposals, for example the Horndeski theories \cite{Horn,kobayashi} of modified gravity, offer several free functions that can only be constrained via observations.

The contrast with our highly-economical proposal could not be sharper.
 The theory we introduce here is more determined even than GR, having one less parameter as $\Lambda$ is turned into a dynamical
field in a way that introduces no additional parameters! 
Not only do we reduce the amount of freedom as compared to GR, but we do so by exploiting ideas which originate in a literature that originally has nothing to do with explaining cosmic acceleration. Rather, the principles which motivate our theory come from developments in quantum gravity.  None have any direct connection to seeking explanations of cosmic acceleration, which is normally the \emph{sole} motivation given for extending the gravitational action
and which ends up either \textit{ad hoc} or instead too general to ever be meaningfully constrained. We hope to alert the community to the possibility of an explanation for dynamics of $\Lambda$ with a deeper theoretical grounding than the usual proposals.

Our main focus in this article is on the simplest `vanilla' form of this theory. As will become apparent, the theory offers a number of routes for extensions, being one example of a small class of theories we have constructed to realize the principle of quasi-topological dynamics. These may also be of considerable phenomenological interest, starting out as they do from a framework substantially differing from the usual GR one. Some of these introduce a single new parameter, bringing the number of parameters back up to the two of general relativity. In the penultimate section we survey the other theories in this class, some of which will be the subject of papers now in progress. 

We note that it is possible that the minimal theory we describe here, while consistent, is too constrained to describe Nature and that one of its variants, having slightly more freedom, may be `just right'. 
But for now our attention is on the most conceptually elegant scenario. For one thing it is the starting point for the whole class. Also it is a consistent diffeomorphism invariant $3+1$ dimensional field theory, simpler than GR but still with local degrees of freedom, and hence an object of some interest in its own right.

This article focuses on the structural and foundational aspects of the theory. Companion papers will develop the cosmology of the model \cite{FRW1,MZ,FRW2}, explore its near cousins, and make a detailed investigation of their viability against Solar System and gravitational wave tests.  Some of the implications for the quantum theory have already been discussed in Refs.~\cite{JL,SJL}.

\section{Melting the cosmological constant}

The work described here resulted from the convergence of several ideas which, when merged together, allowed for the discovery of a natural method for giving dynamics to the cosmological constant in a theory which is more constrained than general relativity. We enumerate these ideas here.  They come from both cosmology and quantum gravity.

Cosmologists are interested in studying modified gravity theories in the infrared
in the hope of understanding dark energy and, perhaps, dark matter.
But most candidates require new fields and new parameters. 
These reduce their testability and explanatory power.
It would therefore be extremely interesting to know if
there is a principle which modifies gravity in a way that gives
dynamics to the dark energy, but has no new parameters or fields.  We will shortly present such
a principle.

Meanwhile, quantum gravity theorists have learned that general relativity and quantum
gravity are in important ways close to topological field 
theories \cite{Plebanski,CDJ,ls-positive}. There are senses in
which the low-energy limit of quantum gravity  is dominated 
by a topological quantum field theory (TQFT).  
One of these describes the gravitational field as arising from defects in a 
TQFT \cite{Bianca, Eugenio}.  Another involves a close connection between
the dynamics of the gravitational field and 
Chern--Simons theories \cite{chopinthesis,Chopin-Lee,Soo2001,Kodama}.
The cosmological 
constant $\Lambda$ plays
an important role in these insights \cite{ls-positive}.  For that reason de Sitter space-time
and associated quantum states are seen in a new light from the perspective of a
constrained or broken TQFT \cite{Kodama}.

This suggests that  any infrared modification of gravity that could have cosmological
implications should be closely tied to topological field theories.  We propose one way to do that, which contemplates that a modified gravity
theory would involve the addition of a term $S^{\rm new}$ to the action governed by the following principle:
\begin{itemize}
\item{ \bf Quasi-topological principle:}   {\it Introduce only new terms in $\Lambda$ 
that are topological when $\Lambda$ is constant. Thus,  $\Lambda$  gets its dynamics from disrupting
a topological invariance.}
\end{itemize}

There are two classes of theories which realize this principle, stemming from each of the two topological invariants available in $3+1$ dimensions which can be formed from a connection alone. These are the even-parity Gauss--Bonnet invariant and the odd-parity Pontryagin invariant. This paper mainly uses the Gauss--Bonnet invariant, with discussion of Pontryagin deferred until Section~\ref{subsec:Pontry}. The basic idea is to add a Gauss--Bonnet term multiplied by a prefactor which is a function of $\Lambda$, which thus becomes purely topological whenever $\Lambda$ is constant. As we will see, the possible forms of this function can be strongly restricted, and in the most elegant version of our theory the function, both in form and in normalization, is uniquely determined.

\subsection{General relativity in first-order form}

In order to establish both notation and the direction of our ideas, we begin by revisiting General Relativity within the form-based first-order Palatini formalism, as described for instance by Wald \cite{Waldbook}.  As is well known, for GR the Palatini formalism yields the same physical theory as the traditional metric-based second-order formalism, but this is not necessarily true for more complex gravitational actions, $f(R)$ theories being an example where the theories are physically distinct.

The importance of this formulation is that the action appears in first-order form,
as a function of both the co-tetrad (frame-field) $e_{\mu}^a$ and the spin connection $\omega_{\mu}^{ab}$, which henceforth will both be expressed as one-forms, $e^{a}$ and $\omega^{ab}$.  The Latin indices live in the internal Lorentzian space-time and the Greek indices in coordinate space-time.  Because we use form notation, we ordinarily suppress space-time indices. We now consider the Palatini action, exploring the obstruction to having a variable cosmological constant $\Lambda$. 
The action is 
\begin{eqnarray}
S^{{\rm GR}} [ e^a, \omega^{ab} , \psi ] =  \frac{1}{8\pi G} \int_{\cal M}  \epsilon_{abcd} \left( e^a \wedge e^b \wedge R^{cd}(\omega) \phantom{\frac{\Lambda}{6} }\right.  \\
\left. -\frac{\Lambda}{6}  e^a \wedge e^b \wedge e^c \wedge e^d 
\right) + S^{\rm matter}[e^a , \psi ]\,.\nonumber
\label{SPal}
\end{eqnarray}
where $\psi $ corresponds to matter fields.\footnote{For simplicity we assume in this article that the matter action depends explicitly only on the frame fields, not the connection, though the generalization (necessary to include spinors) should pose no new technical issues.} Here $R^{cd}$ is the curvature two-form, which in this first-order formalism is defined by 
\f
R^{cd}[\omega]= d \omega^{cd}+\eta_{ab} \omega^{ac} \wedge \omega^{bd}\,,
\label{Riemann}
\ff
where $\eta_{bc} \equiv {\rm diag}(-1,1,1,1)$ is the flat metric with respect to the frame fields.
Note that $R^{cd}$ has two suppressed indices, so in total there are four indices.   The antisymmetric tensor $\epsilon_{abcd}$ in Eq.~\eqref{SPal} contracts the indices so as to extract the curvature scalar as in the usual Einstein--Hilbert action. Note that the forms are already densities so there is no volume term, $\sqrt{-g}$, in the gravitational action.

Our theory can be expressed in other first-order formulations, such as 
Pleba\'nski \cite{Plebanski}, Jacobson--Samuel--Smolin \cite{Sam,tedlee-action}, and the Holst action \cite{Holst}.  But a first-order formulation in which the space-time  connection starts off independent of the metric is necessary, because we will require a field equation for the connection.  Different aspects of the theory may be more accessible from the points of view of various formulations.  We use Palatini here as it is most familiar to cosmologists, who are our main audience.

Since we are in first-order form, we begin with the connection equations of motion.  The torsion two-form $T^a$, is given by the covariant curl of the frame one-form as 
\f
T^a \equiv {\cal D} e^a \,,
\ff
where the covariant derivative is defined as
\f
{\cal D}e^a \equiv de^a + \omega^{a}_{\ \ c}  \wedge e^c \,,
\ff
where $d$ is the exterior derivative. Variation of the action with respect to the connection then yields an equation for the torsion:
\f
0 =  \frac{\delta S^{\rm GR}}{\delta \omega_{ab}}  
 \Longrightarrow {\cal D} \left ( e^a \wedge e^b \right ) =2  T^{[a}\wedge e^{b]}  =0 \,,
\label{eomframefield}
\ff
where, as usual, square brackets indicate antisymmetrization on indices. 

Equation~\eqref{eomframefield} is sufficient to ensure that the torsion $T^a$ identically vanishes; this three-form equation comprises a total of 24 constraints (4 form indices times 6 antisymmetrized `ab' combinations) on the 24 components of the torsion two-form (6 form indices times 4 `a' values). The absence of torsion in GR means that the spin connection is the (torsion-free) Levi-Civita connection, whose coordinate representation as Christoffel symbols ensures that the Palatini action is physically equivalent to the second-order formulation. The usual metric can be obtained from the frame fields via $g_{\mu\nu} =  \eta_{ab} \, e^a_\mu e^b_{\nu}$, where this time we explicitly displayed the form indices.

We look next at the frame-field equation of motion, which yields the Einstein equation\footnote{In this formalism there are 16 Einstein equations, as this is a three-form equation with one additional index, rather than the usual 10. This is because use of frame fields adds an extra redundancy where different frame fields can correspond to the same metric.} 
\begin{eqnarray}
0 =  \frac{\delta S^{\rm GR}}{\delta e^{a}}  
 \Longrightarrow  \hspace*{4.5cm}\nonumber \\
\epsilon_{abcd} e^b\wedge \left ( R^{cd}-\frac{\Lambda}{3}  e^c\wedge e^d \right) 
=  -16\pi G \tau_a \,,
\label{eq2M}
\end{eqnarray}
where the energy--momentum three-form, $ \tau_a$, is given by
\f
 \tau_a \equiv   \frac{1}{2} \frac{\delta S^{\rm matter}}{\delta e^{a}} \,.
 \ff
It is related to the energy--momentum tensor
 \f
 \tilde{T}^{\mu \nu} =\frac{\delta S^{\rm matter}}{\delta g_{\mu \nu}} \,,
 \ff
 by
 \f
 \tilde{T}^{\mu \nu} = \frac{1}{6} 
 \epsilon^{ \alpha \beta \gamma  (\mu}  e_a^{\nu)} \tau_{ \alpha \beta \gamma}^a  \,.
 \ff
 We note is that $ \tilde{T}^{\mu \nu} $ is symmetric by definition (because $g_{\mu \nu}$ is), 
 and is a density of weight one. For ease of notation we also defined the self-dual current
 \f
 {\cal J}^{cd} \equiv R^{cd}-\frac{\Lambda}{3}  e^c\wedge e^d \,,
 \label{SDdef}
 \ff
so the Einstein equations can be written
\f
 \epsilon_{abcd} e^b\wedge  {\cal J}^{cd}
=  -16\pi G \tau_a \,.
\label{eq2M2}
\ff

To gain some insight about how to consistently implement a dynamical cosmological constant, we allow $\Lambda$ to vary and consider the covariant curl of Eq.~\eqref{eq2M}.  Here, the covariant derivative acting on the curvature two-form $R^{ab}$, contains the full connection.  Upon an application of the Bianchi identities ${\cal D} R^{ab} =0$,\footnote{Because our covariant derivative is with respect to the spin connection, it encodes the effects of both curvature and torsion in the Bianchi identities.} we arrive at 
\begin{eqnarray}
\epsilon_{abcd} \left\{ e^b \wedge \left[ \frac{d\Lambda  }{3} \wedge e^c \wedge e^d
+ \frac{2 \Lambda  }{3} T^c \wedge e^d \right]+ T^b \wedge J^{cd} \right\} \nonumber \\
+{\cal D}\tau_a=0 \,. \hspace*{1cm}
\label{consist1}
\end{eqnarray}
For now, let us assume that the energy--momentum tensor is covariantly conserved, ${\cal D}\tau_a=0$, an issue we will return to later in this paper.  Since we already showed above that the torsion vanishes in GR, thereby eliminating the second and third terms of Eq.~\eqref{consist1}, we arrive at the final result that the Bianchi identities imply that
\f
d\Lambda =0 \,,
\ff
hence recovering the known implication that GR only has solutions when $\Lambda$ is a constant. 

This is the point where our work begins: we will seek actions in which the GR equations of motion are unmodified, but despite that admit solutions in which $\Lambda(x,t)$ is allowed to vary, while at the same time satisfying the consistency condition in Eq.~\eqref{consist1}. 

In order to start on this goal we review the properties of self-dual vacuum solutions, as these will motivate the construction that follows. The de Sitter space-time is the
unique solution with $\Lambda > 0$ that has the maximal number of symmetries, or Killing vector fields, but
may also be characterized as the unique space-time with Lorentzian signature
which satisfies the self-dual condition 
\f
R^{ab}[\omega] = \frac{\Lambda  }{3} e^a \wedge e^b \,,
\label{SD1}
\ff
meaning that the self-dual current ${\cal J}^{ab}$ identically vanishes for de Sitter solutions. 
Differentiating Eq.~\eqref{SDdef} also shows that in vacuum ($\tau_a=0$) the covariant derivative of the self-dual current vanishes in GR even if the solution itself is not self-dual.

Equation~\eqref{SD1} defines {\it self-dual solutions with non-vanishing $\Lambda$.}  The form of the Einstein equations \eqref{eq2M} suggest a special role
for solutions of this form, which we see automatically satisfy the vacuum Einstein equations. An important feature is that the self-dual condition implies that the Weyl curvature $C^{abcd}$ vanishes.

Note that we wrote Eq.~\eqref{SD1} using the spin connection $\omega$, which is different from the metric-compatible torsion-free connection.\footnote{Written in terms of the torsion-free connection, $\tilde{\omega}[e]$, Eq.~\eqref{SD1}  becomes second order, since $\tilde{\omega}$ and $e$ are now not independent.} In GR they are same, $\omega=\tilde{\omega}[e]$, but in what follows we'll be interested in solutions with torsion, so the distinction here is very relevant.

\subsection{Dynamical $\Lambda$ from torsion}

Now we can notice something new.  In order to allow for a Bianchi identity consistent varying $\Lambda(x,t)$ we can look at Eq.~\eqref{consist1} and note that, if restricted to the self-dual vacuum solutions, ${\cal J}^{cd}=0$, and in the presence of torsion, it is solved if 
\f
\frac{d\Lambda  }{3} \wedge e^a
+ \frac{2 \Lambda  }{3} T^a=0\,.
\ff
This is strongly suggesting to us to seek a torsion of the form 
\f
T^a = -\frac{d\Lambda}{2 \Lambda} \wedge e^a \,,
\label{T3}
\ff
which will be Bianchi abiding and at the same time allows for a non-constant $\Lambda$ to be included in the Einstein equations.
The torsion Eq.~\eqref{T3} actually ensures that the Bianchi identities \eqref{consist1} are automatically satisfied whatever space-time dependence $\Lambda(x,t)$ has, provided the solution is self-dual.  

So the effect of this torsion is to generate a new infinite set of solutions to the first-order Einstein equations, with the inclusion of torsion in the first-order formalism, given by
\begin{eqnarray}
R^{ab}[\omega ] &=& \frac{\Lambda (x,t) }{3} e^a \wedge e^b \,,
\nonumber
\\
T^a &=& -\frac{d\Lambda}{2 \Lambda} \wedge e^a\,.
\label{SD3}
\end{eqnarray}

We note that these are distinct from the usual de Sitter solution. The torsion $T^a$ is a tensor, so these new solutions are not related to de Sitter space-time by diffeomorphisms and are physically different. We also emphasize that $\Lambda(x,t)$ remains fully unconstrained at this point, and Eqs.~\eqref{SD3} are satisfied by what ever form we choose for $\Lambda$.

In summary, we have found here a new class of solutions which allow for fully space-time varying $\Lambda$ and satisfy the Bianchi Identities. However they cannot be solutions to GR as it does nto support torsion. Now our goal is to find the corresponding theory that does yield such solutions.
 
\section{Finding the consistent action for dynamical $\Lambda$}\label{Fl1}

The new solutions found in Eqs.~\eqref{SD3}, with varying $\Lambda$, are {\it not} solutions of GR and are valid only in a theory for which the torsion has the proposed form. In this section we will derive the form of the action which yields these solutions.

A careful inspection of the action in Eq.~\eqref{SPal} can reveal that, in the absence of matter, it can be made symmetric when exchanging the quantities $e^a\wedge e^b$ and $R^{cd}$ if we add a new term of the form $\epsilon_{abcd} R^{ab}\wedge R^{cd}$. Then in this new form, the action in Eq.~\eqref{SPal} acquires a duality symmetry, and will go back to itself under the swap operation, 
\f
R^{ab} \leftrightarrow \frac{\Lambda}{3} e^a \wedge e^b \,,
\label{duality1}
\ff
The new full gravitational action now becomes
\begin{eqnarray}
S^{{\rm GR+new}} &=& \frac{1}{8\pi G} \int_{\cal M}  \epsilon_{abcd} \left ( e^a \wedge e^b \wedge R^{cd}(\omega) \phantom{\frac{\Lambda}{6}} \right. \\ 
&& \hspace*{-1cm} \left. -\frac{\Lambda}{6}  e^a \wedge e^b \wedge e^c \wedge e^d - \frac{3}{2 \Lambda} R^{ab} (\omega) \wedge R^{cd} (\omega ) \right)\,, \nonumber
\label{Spal-vac}
\end{eqnarray}
and it is now invariant under the duality symmetry given by Eq.~\eqref{duality1}. Note that the same symmetry also fixes the exact form of the coefficient of the new $R\wedge R$ term to be $-3/2\Lambda$. 
By construction this leaves the self-dual condition \eqref{SD1} unchanged. 

Remarkably, we will show that this form of the action is also the form which yields the torsion self-dual solutions Eq.~\eqref{SD3} that we are looking for. 

Additionally we note that the term newly added to the action has the recognizable form of a term well known in the literature, the Gauss--Bonnet term,\footnote{Readers may be more familiar with the traditional tensor representation of the Gauss--Bonnet term
\f 
I^{\rm GB} = - \int_{\cal M} \sqrt{-g} \left(R^2 - 4 R_{\mu\nu}R^{\mu\nu} + R_{\mu\nu\rho\sigma} R^{\mu\nu\rho\sigma} \right)\,.\nonumber
\ff}
\f
I^{\rm GB} =-  \int_{\cal M} \epsilon_{abcd} R^{ab} \wedge R^{cd} \rightarrow \ \ - \int_{\cal M} \frac{3}{2 \Lambda} \epsilon_{abcd} R^{ab} \wedge R^{cd}\,.
\label{disruptGB}
\ff
The Gauss--Bonnet term in the first expression in Eq.~\eqref{disruptGB} is a topological invariant and hence has no effect in the equations of motion of the corresponding action.    However, in the second expression we have disrupted this term by the inclusion of the coefficient $-3/2\Lambda$, which is required by the duality symmetry of Eq.~\eqref{duality1}. We call the resulting term quasi-topological Gauss--Bonnet. These couplings are analogous to axion couplings to topological invariants in Yang--Mills theory \cite{Weinberg:1977ma,Wilczek:1977pj}.

The Gauss--Bonnet term has featured prominently in both the gravitational and cosmological literatures. The consequences of adding a Gauss--Bonnet term to the action, but with constant coefficients,  have been explored in Ref.~\cite{ABJ}.  There may be also a relationship to the extended Chern--Simons theory of Ref.~\cite{J-Pi}, with the arbitrary one-form field $v_a$ taken as proportional to $d\Lambda$. There is an enormous cosmological literature exploiting the Gauss--Bonnet term in various guises, to which Ref.~\cite{modifiedg-review} provides an entry point. Typically this involves adding a function of it to the Einstein--Hilbert action, for example Ref.~\cite{fgb}, or using a string-inspired coupling to a scalar field, as in Ref.~\cite{scalGB}. Actions with a function of a scalar field multiplying the Gauss--Bonnet term are known to lie within the Horndeski class of theories \cite{kob2,kobayashi}; this is certainly not obvious at first sight, but nor is it of direct relevance to us as Horndeski theories are typically analysed in the physically-inequivalent second-order form. For exceptions that do use the first-order form, though not with our action, see Refs.~\cite{TolZal,BDGL}, the latter analysing a teleparallel version of Horndeski with torsion. An extremely general action, encompassing ours as a special case, was explored in first-order form in the different context of cosmological signature change in Ref.~\cite{MRWZ}.  

Notice also that our action Eq.~\eqref{Spal-vac}, but with constant $\Lambda$, is reminiscent of the 
MacDowell--Mansouri formulation of GR \cite{MM,ArtemLee,laurent,sa}, 
which is based on a broken five-dimensional Gauss--Bonnet term with a $2/3\Lambda$ prefactor. 
A 4+1 split of the space-time then yields three terms of the same form as our action.\footnote{We thank Latham Boyle for pointing out this connection to us.} This also connects it to the quadratic Lovelock-Unique-Vacuum (LUV) theories (e.g.\ Ref.~\cite{LUV}) where again only constant $\Lambda$ has previously been considered.

The Einstein equation, the equation of motion for the frame-field $e^a$ given by Eq.~\eqref{eq2M}, is unchanged by the addition of the quasi-topological term, but the connection equation of motion in Eq.~\eqref{eomframefield} is now:
\f
0 =  \frac{\delta S^{\rm GR+new}}{\delta \omega_{ab}}  
 \rightarrow
S^{ab}\equiv  T^{[a}\wedge e^{b]}  =
- \frac{3}{2\Lambda^2}d\Lambda\wedge R^{ab}\,.
\label{eq6M}
\ff
By the same counting argument we used for GR, $S^{ab}$ encodes the same information as the torsion tensor itself. Inverting this via a calculation
lets us find the torsion tensor: 
\f T^{a}_{\mu\nu} = \frac{3}{2} \epsilon^{\alpha\beta}_{\;\;\mu\nu}e^{\sigma}_{b}\partial_{\alpha}({\rm ln}\Lambda)R^{ab}_{\beta\sigma} \,.
\label{T17}
\ff

We see that the torsion becomes a function of the curvature two-form which will necessarily lead to a modified Einstein equation that is non-linear in the full connection.  Indeed Eq.~(\ref{T17})
is a quadratic equation for the components of $T^{a}_{\mu\nu}$, because the full curvature
$R^{ab}_{\beta\sigma}(\omega )$ is itself a quadratic function of the components of torsion.

It is important to note that this non-linear character of the connection field equation, which determines the torsion tensor, still ensures the consistency of the Einstein equation, through the vanishing of its covariant curl, Eq.~(\ref{consist1}).\footnote{The issues of covariant conservation
and the consistency of the Einstein equations in the presence of torsion are
discussed in Refs.~\cite{Trautman,kibble,Hehl}.} We note for future applications that when gravity is coupled to fermions the issue of covariant conservation will be more subtle.  The fermionic energy--momentum tensor is connection dependent and will carry anti-symmetric components of energy--momentum, yielding extra sources of spin currents. These have been considered by many authors, for example Refs.~\cite{kibble,Hehl}.  We will investigate our theory in the context of spin-currents and fermionic matter in a forthcoming paper.

Our system is analytically tractable if we restrict to the self-dual sector. Under the duality operation in Eq.~\eqref{duality1}, the connection equation in Eq.~\eqref{eq6M} can be rewritten as
\f
S^{ab}= T^{[a}\wedge e^{b]}  =
- \frac{d \Lambda }{2\Lambda} \wedge e^a \wedge e^b \,.
\label{eq7M}
\ff
This confirms that within the self-dual sector our action generates the desired torsion, and hence allows the solutions with arbitrary space-time variation of $\Lambda(x,t)$ displayed in Eq.~\eqref{SD3}.

\section{The $\Lambda$ equation of motion}

At this point we have defined a new set of vacuum solutions, given by Eq.~\eqref{SD3}, to the modified action Eq.~\eqref{Spal-vac}, which allow for variation of $\Lambda$ via the inclusion of torsion while satisfying the Bianchi Identities. With an eye towards moving beyond the self-dual category, we introduce a further novel step which is to obtain an additional equation of motion by varying the action with respect to $\Lambda$.\footnote{A different proposal for allowing $\Lambda$ to be varied in the action, based on quantum partition functions, has been made by Barrow and Shaw \cite{barrowshaw}, though in their case $\Lambda$ remains a space-time constant.} As we noted in the introduction, such a variation cannot be carried out in GR as it yields physically useless results, so there $\Lambda$ is left as a parameter to be determined from observations. But now, having observed that $\Lambda$ can be a function of location, it is appropriate to also include its variation in the action. This gives the new equation
\f
0 =  \frac{\delta S^{\rm GR+new}}{\delta \Lambda}  
 \Longrightarrow
\frac{\Lambda^2}{9}=
\frac{\epsilon_{abcd} R^{ab} \wedge R^{cd}}{e^4} \,.
\label{eq8M}
\ff
where $e^4 \equiv \epsilon_{abcd} \left(e^a \wedge e^b \wedge e^c \wedge e^d\right)$ is the volume four-form. This equation provides an algebraic equation for $\Lambda$.

We first note that within the self-dual sector this equation is automatically satisfied, simply stating $\Lambda^2 = \Lambda^2$. This was inevitable; as we have already shown that under the self-dual condition an arbitrary function $\Lambda(x,t)$ is a solution, it can't be subject to any additional constraint equation.

Beyond self-duality, however, the implications of Eq.~\eqref{eq8M} are dramatic and two-fold:
\begin{itemize}
\item The equation of motion for $\Lambda(x,t)$ in Eq.~\eqref{eq8M} comes at no extra cost. We have not had the need to introduce extra degrees of freedom, such as scalar fields or kinetic terms, to obtain it.
\item Equation~\eqref{eq8M} fixes the value of $\Lambda(x,t)$ to the Gauss--Bonnet density, so $\Lambda$ is now determined within the theory, rather than being a free quantity that has to be measured. This implies that we have just reduced, by one, the number of free parameters in our theory, compared to number of degrees of freedom in GR+$\Lambda$.
\end{itemize}
These are the key points of this work, which distinguish it from existing literature. 

\section{Solutions without self-duality}

While the self-dual solutions have played a central role in motivating the construction of our theory, to be physically interesting it must be able to describe general situations where the self-dual assumption is lifted. In particular loss of self-duality is inevitable once matter is included, as well as in the vacuum case if there is Weyl curvature as for instance in the Schwarzschild solution.

Since our theory is defined by an action principle, we should expect its equations of motion to be consistent.  It appears to have local degrees of freedom and so is not a topological field theory.  But, at the same time, it is also clear that it is not general relativity. The new theory has a variable $\Lambda$, which satisfies its own field equation, which is unlike anything in general relativity.  As we saw, like general relativity in first-order form,  the connection and in particular the torsion are determined by a non-linear extension of the connection field equations.
  
In particular, we have to answer the important question of whether there are solutions beyond the self-dual sector to the full theory defined by Eq.~\eqref{Spal-vac}, both in the absence or presence of matter.  This is the subject of several lines of ongoing investigation. Here is what is known at present:

\begin{itemize}

\item{}  There are non-trivial solutions, with matter, in the reduction to homogeneous
and isotropic solutions.  These generalized FRW solutions are of some interest, as $\Lambda (t)$ becomes locked to the matter density in a manner reminiscent of quintessence scaling solutions, and are the subject of separate papers \cite{FRW1,MZ}.

\item{} In the homogeneous and isotropic case, without matter, all solutions satisfy
the self-dual condition with arbitrary $\Lambda$, as shown also in Ref.~\cite{FRW1}.

\item{}There is an argument, based on the Pleba\'nski formulation that in the Euclidean vacuum case, all solutions are in the self-dual 
sector \cite{Kirill}. In the Lorentzian case, this argument fails to show all solutions are self-dual,
and in fact suggests a strategy to construct more solutions, which have non-vanishing but null
Weyl tensor \cite{WolfgangLee} 
\f
C^2 = C_{abcd} C^{abcd} =0 \,.
\ff
\end{itemize}

\section{Options for varying $\Lambda$}

Here we pause to reflect on what it means for $\Lambda$ to be variable in our context. Essentially there are two possibilities; either the action is extremized under variations with respect to $\Lambda$, as in our proposal, or it is not, as in GR. As the variation imposes extra dynamical restrictions on the model, it yields a more predictive framework, with potentially fewer parameters than GR+$\Lambda$.

We have found that this variation gives three possible scenarios, depending on the physical circumstance, and examples of each are presently under investigation.
\begin{itemize}

\item[A.] The $\Lambda$ equations of motion are redundant, and so impose no new 
constraints.  The choice of the function $\Lambda (x^\mu )$  is then free and 
unconstrained. We saw this occur here in the self-dual sector.

\item[B.] The new equation of motion is a constraint which ties $\Lambda$  to be a fixed function of the matter density and/or Weyl curvature. We expect this to be the generic situation in physically-realistic scenarios of our vanilla theory. A cosmological example of this behaviour is described in Ref.~\cite{FRW1}.

\item[C.] The other equations of motion induce a kinetic energy term for  $\Lambda$, giving it dynamics similar to a conventional field.  An example of this is described in Section~\ref{subsec:kinetic} below.  The implications for cosmology are developed in Ref.~\cite{FRW2}.
\end{itemize}

We note an alternative possible outcome of variation w.r.t.\ $\Lambda$, not realised in our theory, is for $\Lambda$ to be a single number resulting from extremizing some functional over all of space-time, as in the sequestering scenario of Ref.~\cite{KP}. 

Were we instead to choose not to vary the action w.r.t\ $\Lambda$, we would have considerable additional freedom to specify it. For instance, it could be a given function of space-time, set ultimately by deeper considerations coming from cosmology or quantum theory, but arbitrary at this level.  Alternatively it could be 
 constant on some dynamically-preferred three-slicing of space-time,
making it a function of a time parameter that labels the slices, as described in Ref.~\cite{SJL}.

\section{Extensions to the vanilla theory}

We have so far kept the focus on the simplest form of the theory, due to its conceptual elegance and novel properties. In particular, the additional quasi-Gauss--Bonnet term in the action is precisely specified adding no new parameter or functional degrees to the action, and indeed potentially removing them by permitting self-consistent variation of the action with respect to $\Lambda$. However the theory is also of interest as a different starting point from GR to make extensions to the gravitational theory, as might be guided by current and future observational constraints. In this short section we outline several such extensions, for more detailed exploration in future work.

\subsection{Generalising the quasi-topological term}

The simplest possible extension to our theory is to let the new quasi-Gauss--Bonnet term be controlled by a constant multiplier $\theta$, while retaining its functional dependence on $1/\Lambda$. The action now reads
\begin{eqnarray}
S^{\theta} & = & \frac{1}{8\pi G} \int_{\cal M}  \epsilon_{abcd} \left ( e^a \wedge e^b \wedge R^{cd}(\omega) \phantom{\frac{\Lambda}{6}} \right. \\
&& \hspace*{-1cm} \left.-\frac{\Lambda}{6}  e^a \wedge e^b \wedge e^c \wedge e^d - \frac{3\theta}{2 \Lambda} R^{ab} (\omega) \wedge R^{cd} (\omega ) \right) + S^{\rm matter}\,. \nonumber
\label{Stheta}
\end{eqnarray}
where $\theta = 0$ returns us to GR (provided we make a simultaneous decision not to vary the action w.r.t.\ $\Lambda$), 
and $\theta = 1$ is the special theory we have been discussing thus far. The Einstein equation remains unchanged, while the $\Lambda$ equation of motion now reads
\f
0 =  \frac{\delta S^{\theta}}{\delta \Lambda}  
 \rightarrow
\frac{\Lambda^2}{9}=
\frac{\theta \epsilon_{abcd} R^{ab} \wedge R^{cd}}{e^4} \,,
\label{eq11M}
\ff
and the connection equation Eq.~(\ref{eq6M}) also acquires a $\theta$ multiplier on its right-hand side.

Introducing the $\theta$ parameter takes away some of the elegance of the original construction, which is why we have mostly focussed on the vanilla theory in this article. In particular self-dual solutions no longer exist, since the equation of motion immediately implies $\Lambda^2 = \theta \Lambda^2$ in that case. In a sense, $\theta-1$ measures the extent to which the duality symmetry is broken. However, our use of self-dual solutions is primarily motivatory, since anyway cases of physical interest will not obey that condition. As in the $\theta = 1$ case, Eq.~\eqref{eq11M} will fix $\Lambda$ in terms of the other variables describing a chosen situation.

There is the further disadvantage of introducing an extra free parameter, but even then this is only returning us to the level of freedom of GR+$\Lambda$, by swapping in $\theta$ in place of $\Lambda$ as an undetermined parameter. Since the theory remains radically different from GR it surely merits careful study as to its self-consistency and observational viability. This additional freedom does increase the opportunity to match observations, particularly in the construction of viable cosmological models which we study in Refs.~\cite{FRW1,MZ}.

\subsection{Using the Pontryagin invariant}
\label{subsec:Pontry}

An alternative to our Gauss--Bonnet theory would be to add, in the place of Eq.~\eqref{disruptGB}, a similarly-modified term involving the Pontryagin invariant \cite{JL,SJL}: 
\f
S^{\rm new-Pont}= 
- \frac{1}{8 \pi G }   \int_{\cal M}  \frac{3}{2 \Lambda}  R^{ab} \wedge R_{ab}\,.
\label{new1}
\ff
This modification breaks parity, and connects with a class of parity-odd
modifications of GR studied in Ref.~\cite{Alexander:2009tp}. It leads to interesting new complex solutions, which may play a role in path-integral approaches to the quantum theory.

It does inspire a very interesting hypothesis about a physical link between the (dynamical) cosmological constant and a gravitational chiral anomaly \cite{SJL}.  
The latter is related to the difference between left- and right-handed particle creation rates \cite{ChrisDuff}: 
\f
{\cal D}_\mu J^\mu_5 = \frac{3}{16\pi^2} R^{ab} \wedge R_{ab}\,.
\ff
We can use the $\Lambda$ equation of motion to obtain
\f
\frac{\Lambda^2}{9}= \frac{16\pi^2}{3} \, \frac{{\cal D}_\mu J_5^\mu }{\sqrt{-g}} \,,
\ff
which posits a relationship between $\Lambda$ and a particle (mass) scale, given by the empirical relation 
\f
G \Lambda \approx (\Delta m_\nu )^4\,,
\ff
where $\Delta m_\nu \approx 3 \times 10^{-3}\,{\rm eV}$. This reminds us of the often-remarked  coincidence of the observed neutrino and dark energy mass scales.

\subsection{Giving $\Lambda$ a kinetic energy from torsion-squared terms}
\label{subsec:kinetic}

In the original theory, $\Lambda$ is dynamical in the sense that its value is determined by a field equation which results from varying it in the action.  But since there are no derivatives in that field equation, there is no propagating mode.  The result, as we show in detail in the generalized FRW case \cite{FRW1}, is that the dark energy is locked to the matter density.  This has interesting implications for cosmology, as we
discuss there.

We can instead give the cosmological constant a propagating mode by adding a term to the
action proportional to the square of the torsion, such as
\f
S^{T^2}=\alpha  \int \sqrt{-g} \, g^{\mu \nu}  T_{\mu \alpha}^{\ \ \ \alpha} T_{\nu \beta}^{\ \ \ \beta} \,.
\label{ST2}
\ff
As the torsion is a tensor this is consistent with diffeomorphism invariance.  If we add this term by hand, the cost is a single new dimensionless parameter, $\alpha$,
which brings the number of parameters back up to the two of general
relativity.  But there are good reasons to expect that such a term is anyway induced by
quantum corrections.  Another mechanism that will induce such a term
would be a fermion condensate, of the kind described in Ref.~\cite{Alexander:2009uu}.

Either way, adding the term in Eq.~\eqref{ST2} to the action gives us a standard kinetic energy for $\ln \Lambda$. Now $\Lambda$ has a field equation that does not restrict vacuum solutions to the self-dual sector. 

If the self-dual conditions in Eq.~\eqref{SD1} are satisfied, then a calculation shows that Eq.~\eqref{T3} gives
\f
S^{T^2}= \frac{3\alpha}{4}  \int \sqrt{-g} \, g^{\alpha \beta } \partial_\alpha \ln \Lambda \partial_\beta \ln\Lambda\, .
\label{ST2-2}
\ff
The result is then that $\Lambda$ has a propagating mode.  This has important implications,
which will be the  subject of another paper \cite{LKE}.

\section{Conclusion}

In this paper we have introduced a class of modifications of general relativity, which feature a varying cosmological constant. These are consequences of a new dynamical principle we introduce, called the {\it quasi-topological principle.} The principle can be applied to two topological invariants, the Gauss--Bonnet and the Pontryagin invariants, giving rise to two parallel sets of theories.  In this paper we focused mostly on the Gauss--Bonnet extension, and have seen how torsion can enable a consistent space-time variation of $\Lambda$.

Because the cosmological constant becomes variable and dynamic in the sense that a new field equation arises from its variation, no new parameter emerges.  However these new field equations are algebraic and do not allow the emergence of a propagating mode associated with the variations of $\Lambda$.  Instead, $\Lambda$ molds itself to the matter distribution.  In companion articles we present in detail the implications for cosmology \cite{FRW1,MZ}.  There are also possible implications for dark matter through the expected enhancement of $\Lambda$ in the vicinity of normal matter.  In the Gauss--Bonnet case, the theory shows potential to explain the coincidence problem, as a consequence of the molding of $\Lambda$ to the matter density.  However, we anticipate challenges to fit the whole history of our universe, particularly during the radiation-dominated regime.  

The theory has various generalizations. One is to move the coefficient of the quasi-topological
term away from the magical $3/2\Lambda$ value, motivated by the duality symmetry, to the more general $3\theta/2\Lambda$. Our preliminary studies show that the theory predicts very different behaviour depending whether $\theta$ is set equal to unity or not.
Alternatively, using the Pontryagin invariant in place of Gauss--Bonnet gives rise to a precise relationship between $\Lambda$ and a left--right asymmetry in particle production rates, mediated by a gravitational anomaly \cite{SJL}.  There are two natural extensions of these models, in which a kinetic energy term naturally arises for the logarithm of $\Lambda$, giving the dark energy a more conventional dynamics via the emergence of propagating modes.  This dynamics is also studied  in the FRW case in a forthcoming paper \cite{FRW2}.   Finally, we can simply add a torsion-squared term to the action to induce a kinetic energy term for $\Lambda$.   Each of these theories has interesting implications for cosmology.

\begin{acknowledgments}

We thank Latham Boyle, Robert Brandenberger, Laurent Freidel, Jim Gates, Marc Henneaux, Chris Hull, Ted Jacobson, David Jennings,  Robert Mann, Evan McDonough, Emil Mottola, Roger Penrose, Sumati Surya, Francesca Vidotto, Yigit Yarig, and Tom Zlo\'snik for discussions and correspondence.   We are especially grateful to Kirill Krasnov and Wolfgang Wieland for crucial insights, and to Jens Boos for advice on working with spin--torsion theory.  

This research was supported in part by Perimeter Institute for Theoretical Physics. Research at Perimeter Institute is supported by the Government of Canada through Industry Canada and by the Province of Ontario through the Ministry of Research and Innovation. This research was also partly supported by grants from NSERC and FQXi. M.C.\ was supported by Funda\c{c}\~{a}o para a Ci\^{e}ncia e a Tecnologia (FCT) through grant SFRH/BPD/111010/2015 (Portugal), and J.M.\ by a Consolidated STFC grant. M.C., A.R.L., and L.S.\ are especially thankful to the John Templeton Foundation for their generous support of this project. 

\end{acknowledgments}

\end{document}